\documentclass[a4paper,11pt]{article}
\usepackage{pos}

\usepackage[english]{babel}
\usepackage[utf8]{inputenc}

\usepackage{lineno,hyperref}

\usepackage{dcolumn}
\usepackage{placeins} 
\usepackage{ragged2e}
\usepackage{etoolbox}
\usepackage{bm}
\usepackage{lipsum}

\usepackage{amsmath}
\usepackage{graphicx,epsfig,wrapfig}
\usepackage{subcaption}
\usepackage[compatibility=false]{caption}

\usepackage{makeidx}
\usepackage{amsfonts}
\usepackage{amsmath}
\usepackage{mathtools}
\usepackage{fixmath}
\usepackage[capitalize]{cleveref}
\usepackage{slashed}
\usepackage{bm}
\usepackage{multirow}

\usepackage{booktabs}
\usepackage{adjustbox}
\usepackage{comment}
\usepackage{dsfont}
\usepackage{float}

\usepackage{color}
\usepackage[dvipsnames]{xcolor}
\usepackage[normalem]{ulem}

\definecolor{myGreen}{rgb}{0.2,0.72,0.2}
\definecolor{myGold}{rgb}{0.83,0.69,0.22}
\definecolor{darkgreen}{RGB}{0,120,0}
\renewcommand\sout{\bgroup \color[rgb]{0.55,0.00,0.99} \ULdepth=-.5ex \ULset}

\title{Matching relations for gluon TMDs up to one-loop accuracy}

\author*[a,b]{Alessio Carmelo Alvaro}
\author[c,d]{Nanako Kato}
\author[a,b]{Barbara Pasquini}
\author[c,d]{Cristian Pisano}
\author[a,b]{Simone Rodini}

\affiliation[a]{Dipartimento di Fisica ``A. Volta", Universit\`a degli Studi di Pavia, \\
I-27100 Pavia, Italy}

\affiliation[b]{Istituto Nazionale di Fisica Nucleare, Sezione di Pavia, \\
I-27100 Pavia, Italy}

\affiliation[c]{Dipartimento di Fisica, Universit\`a degli Studi di Cagliari, \\
Cittadella Universitaria, I-09042 Monserrato (CA), Italy}
\affiliation[d]{Istituto Nazionale di Fisica Nucleare, Sezione di Cagliari, \\ Cittadella Universitaria, I-09042 Monserrato (CA), Italy}

\emailAdd{alessiocarmelo.alvaro01@universitadipavia.it}
\emailAdd{nanako.kato@dsf.unica.it}
\emailAdd{barbara.pasquini@unipv.it}
\emailAdd{cristian.pisano@unica.it}
\emailAdd{simone.rodini@unipv.it}

\abstract{We present new results on the matching relations for gluon transverse momentum dependent parton distribution functions up to one-loop accuracy. 
At tree-level, employing the spinor formalism, we obtain these relations up to twist 3 accuracy for both T-even and T-odd distributions, including the hadron-mass corrections. 
At one-loop order, we extend the parton-in-parton approach to include higher-twist operators in the small-$b$ expansion. 
As a result, we derive for the first time the Wandzura-Wilczek relation for the worm-gear T distributions and establish a systematic method to include mass corrections at one-loop and twist-2 accuracies. }

\FullConference{
The 33rd International Workshop on Deep Inelastic Scattering and Related Subjects (DIS2026)\\
4 - 8 May 2026\\
Bologna, Italy\\
}

\begin{document}
\maketitle

\section{Introduction}
\label{sec:introduction}

Over the last thirty years the investigation of the three dimensional momentum structure of the proton has gained increasing relevance. 
Traditional collinear parton distribution functions (PDFs) have been generalized to include all the three components of the parton momentum through transverse momentum dependent (TMD) PDFs (hereafter, TMDs). 
Despite significant effort to provide a complete picture of the properties of these distributions, particularly regarding their rigorous definition and evolution equations, reliable  extractions are available only for the unpolarized 
(see~\cite{Lorce:2025aqp} and references therein).
In the gluon sector, progress has been slowed by the scarcity of both theoretical constraints and experimental data, limiting current analyses mainly to unpolarized hadrons~\cite{Boer:2010zf,Boer:2012bt,Boer:2011kf,Kato:2024vzt,Lansberg:2017dzg}.

Although TMDs are non perturbative objects and therefore cannot be deduced from first principles, their functional form can be theoretically constrained in specific kinematic regimes.
Working in $b$-space (where $b$ is the Fourier conjugate variable of the transverse momentum of the parton), a generic TMD operator $\Phi$, referred to a parton $i=q,g$ with polarization fixed by the Dirac matrix structure $\Gamma$ and carrying a fraction $x$ of the light-cone hadron momentum, can be perturbatively expanded in powers of the strong coupling $\alpha_{\rm s}$  as~\cite{Bertone:2025vgy,Rein:2022odl}
\begin{equation}
\label{eq:all-order-matching}
    \Phi_{i}^{[\Gamma]} (x,b) = \sum\nolimits_n \alpha_{\rm s}^n \sum\nolimits_j \int_0^1 du \int dy \, \delta(x-uy) \, \mathcal{C}_{ij}^{[\Gamma/\Lambda],(n)}(u,y,b) \, \phi_j(y) + O(b^2),
\end{equation}
where $\phi$ is a collinear operator and $\mathcal{C}$ is the so-called matching coefficient.
At the distribution level, Eq.~\eqref{eq:all-order-matching} translates to relations between TMDs and collinear PDFs, known as matching relations. The
$O(b^2)$ terms are usually neglected beyond tree-level. These terms give rise to hadron-mass corrections to the matching relations~\cite{Moos:2020wvd,Rodini:2023mnh}.
In this work we provide their expressions up to next-to-leading order (NLO) accuracy. 
For all the necessary definitions and details on the calculation, we refer the reader to our work~\cite{Alvaro:2026nip}.

\section{Tree-level computation}
\label{sec:tree-level}

At tree-level, the quantum fields can be treated as classical, allowing for a Taylor expansion around $b=0$ of the gluon TMD correlator in terms of collinear operators. This leads to
\begin{equation}
\label{eq:OPE-tree-level}
    G^{\mu\nu} (x,b) = \sum_{n=0}^\infty \frac{1}{n!} b_{\mu_1} \dots b_{\mu_n} \, \left( \partial^{\mu_1}_T \dots \partial^{\mu_n}_T \, G^{\mu\nu}(x,b) \right)|_{b=0} \, .
\end{equation}
Each collinear operator in the series is a linear combination of operators with geometrical twists $t$ such that $2\le t \le n+2$. 
In Ref.~\cite{Moos:2020wvd} 
a strategy grounded in the spinor formalism was developed to extract a specific twist component 
(twist-decomposition) 
from a generic operator in the series~\eqref{eq:OPE-tree-level}.
Since the Lorentz group $SO(3,1)$ is locally  isomorphic to the group of complex unimodular matrices $SL(2,\mathbb{C})$, each four vector can be mapped into a hermitian matrix.
Within this formalism, the twist decomposition translates into the application of proper differential operators.

At the current stage, the twist decomposition is usually truncated at the geometrical twist 3 components, as collinear PDFs beyond twist 3 are essentially unknown and, in any case,  not relevant for current phenomenology.
The computation starts from the TMD correlator written in position space and follows these steps: 
(i) compactification of the operator, by restricting the Wilson lines to a finite length $L$; 
(ii) expansion of the operator around $b=0$, followed by an expansion of the fields around   $z_i=L$ (where $z_i$ are the positions of the fields); 
(iii) translation of the whole expression into the spinor formalism; 
(iv) extraction of a specific twist component through  the application of appropriate derivative operators;
(v) evaluation of the resulting expression between  hadron states with the same momentum; 
(vi) taking the limit $L\to\mp\infty$ where the sign of the infinity is determined by the specific process (this step is responsible for the sign change in T-odd distributions);
(vii) Fourier transform to $x$ space.
Following this systematic procedure allows us to obtain the tree-level matching relations for all leading-twist gluon TMDs.

We summarize our results in Tab.~\ref{tab:summary}. The full set of the mass series can be found in Ref.~\cite{Alvaro:2026nip}. 
The relevant new results are the small-$b$ expansion for all the T-odd distributions $f_{1T}^\perp$, $h_{1T}$ and $h_{1L}^\perp$.
For the Sivers and transversity distributions the leading term is complete, while the worm-gear L distribution requires the inclusion of the twist-4 component to be finalized. 
The pretzelosity $h_{1T}^\perp$ has no matching up to twist 3 but is expected to match onto twist-4 and twist-5 collinear PDFs.
Finally, we have derived the twist 3 contribution to the worm-gear T distribution $g_{1T}$, which is given by a complicated convolution of quark and gluon twist 3 PDFs.

As an example, we report the result of the unpolarized distribution
\begin{equation}
\label{eq:tree-level-unpolarized}
    f_1^g(x,b) = f_g(x) + \sum_{k=1}^\infty \frac{1}{k!(k-1)!} \left( \frac{x^2M^2b^2}{4} \right)^k \int_0^1 du \int dy \, \delta(x-uy) \left( \frac{\bar{u}}{u} \right)^{k-1} \, f_g(y) \, .
\end{equation}
The leading term in Eq.~\eqref{eq:tree-level-unpolarized} is very well known and reflects the fact that the unpolarized TMD reduces to its collinear counterpart for small values of $b$. 
The second contribution is  the mass series.  Because  $b^2<0$, we have an alternating series with a double-factorial suppression, closely resembling the Taylor expansion of a Bessel function of the first kind. 
The fact that this series  can be expressed in closed form is of fundamental importance for numerical implementations.

\section{One-loop computation}
\label{sec:one-loop}
The one-loop matching coefficients are obtained within the parton-in-parton framework~\cite{Bertone:2025vgy}.
In this approach, 
computing the coefficients translates into analyzing appropriate matrix elements between massless, on-shell, (un)polarized partonic states, which can be perturbatively calculated using standard Feynman rules and momentum-space techniques. 
The different polarizations of the final quark and gluon states are taken into account by employing suitable projection operators.

We provide an extension of this method that allows us to include the neglected terms of order $b^2$ in Eq.~\eqref{eq:all-order-matching}. The core idea is to consider external partonic states with a non-zero  transverse momentum, $p_T\ne0$. 
From a practical point of view, this modification does not increase the complexity of the calculation, except for a minor adjustment in the projectors onto the final gluon states.
As a final result of our computation, we obtain the matching coefficients in the "extended" parton-in-parton framework. 
They have the form $\mathcal{C}_{\text{ext}}=\mathcal{C}_{\text{std}} e^{iu(b\cdot p_T)}$, where $\mathcal{C}_{\text{std}}$ are the coefficients obtained with the standard parton-in-parton calculations. Their expressions can be found, e.g., in~\cite{Bertone:2025vgy,Gutierrez-Reyes:2017glx}.
In position space, the transverse momentum translates into the derivative operator $p_T^\mu\to -i \partial^\mu_T$. This implies that the extra phase $e^{iu(b \cdot p_T)}$ acts as the generating function of the Taylor series around $b=0$. 
Through this procedure, we are able to extend Eq.~\eqref{eq:all-order-matching}  at one-loop order to include all  terms in the $b$ expansion:
\begin{equation}
\begin{split}
    \Phi_{i}^{[\Gamma]} (x,b) =& \alpha_{\rm s} \sum\nolimits_j \int_0^1 du \int dy \delta(x-uy) \mathcal{C}_{ij}^{[\Gamma/\Lambda],(1)}(u,y,b) \\
    & \times \sum_{n=0}^\infty \frac{u^n}{n!} b_{\mu_1} \dots b_{\mu_n} \left[ \partial_T^{\mu_1}\dots \partial_T^{\mu_n} \phi_j^{[\Lambda]}(y,b) \right]_{|b=0} \, .
\end{split}
\end{equation}
We can now substitute  the second line with the twist decomposition of the quark~\cite{Moos:2020wvd} or gluon~\cite{Alvaro:2026nip} TMD correlators to find the matching relations for the gluon distributions onto quark and gluon twist 2 PDFs. 
These procedure yields the two main results of our one-loop computation: the Wandzura-Wilczek relation of the worm-gear T distribution and the inclusion of all the mass corrections. 
At the current stage, we truncate the expansion at twist 2, as the inclusion of twist 3 collinear PDFs would require the evaluation of Feynman diagrams with three external partons.

\begin{table}[t]
\centering
\begin{tabular}{c|c|c|c|c}
    Distribution & Twist of leading matching & Tw2 & Tw3 & Accuracy \\
\hline
    $f_1^g$ & Tw2 & $f_g \, , f_1$ & - & N\textsuperscript{3}LO \\
    $h_1^{\perp g}$ & Tw2 & $f_g \, , f_1$ & - & N\textsuperscript{3}LO \\
    $g_{1L}^g$ & Tw2 & $\Delta f_g \, , g_1$ & -  & N\textsuperscript{3}LO \\
    $g_{1T}^g$ & Tw2-3 & $\Delta f_g \, , g_1$  & $\mathcal{F} \, , \mathcal{T}$   & NLO/LO  \\
\hline 
    $f_{1T}^{\perp g}$ & Tw3 & - & $2F_2^++F_4^+$  & LO \\
    $h_{1T}^g$ & Tw3 & - & $2F_2^+-2F_4^+$  & LO \\ 
    $h_{1L}^{\perp g}$ & Tw3-4 & -  & $2F_2^+-2F_4^+$  & LO \\
    $h_{1T}^{\perp g}$ & Tw4-5 & -  & -  & LO \\
\end{tabular}
    \caption{Summary of existing results for the matching relations of gluon TMDs, including our findings and higher-loop computations from Refs.~\cite{Ebert:2020yqt,Zhu:2025gts,Zhu:2025ixc}.
    The upper (lower) part of the table corresponds to T-even (T-odd) distributions. Dashes denote cases where matching onto PDFs is absent. Double entries in the last column indicate different accuracies for twist 2 and 3 matching.}
    \label{tab:summary}
\end{table}

\section{Conclusions}
\label{sec:conclusions}

In this work we derived the matching relations up to one-loop accuracy for gluon TMDs, including all  hadron mass corrections. The results are summarized in Tab.~\ref{tab:summary}.
At tree-level, we considered both twist 2 and 3 PDFs in our analysis, providing the small-$b$ expansion for T-odd distributions as well as the higher twist component of the worm-gear T.
At one-loop order, we obtained the Wandzura-Wilczek approximation of the worm-gear T.
These results represent a significant improvement in our knowledge of the functional form of gluon TMDs, in particular for polarized distributions,
offering crucial inputs for the phenomenology of polarized gluon observables at current and next-generation colliders.
From a methodological perspective, the main result consists in the extension of the parton-in-parton approach.
This extended approach combines the computational simplicity  of the standard approach with the capability to include all the small-$b$ series. 
With minor adjustments, this new methodology can be generalized to a wide range of applications, including the matching relations for leading- and next-to-leading-power quark TMDs, higher-loop calculations, and the small-$b$ expansion of generalized TMD operators.


\end{document}